\documentclass[]{interact}

\usepackage{epstopdf}
\usepackage{subfig}

\usepackage{natbib}
\usepackage{graphicx}
\usepackage{placeins}
\usepackage{array}
\usepackage[font=scriptsize,labelfont=bf,labelsep=space]{caption}
\usepackage[flushleft]{threeparttable}
\usepackage[hyphens,spaces,obeyspaces]{url}
\usepackage{tikz}
\usetikzlibrary{arrows,shapes,calc}
\usepackage{pgf}
\usepackage{comment}
\usepackage{booktabs}

\theoremstyle{plain}

\theoremstyle{definition}

\theoremstyle{remark}

\tikzstyle{block} = [rectangle, draw, fill=white,
text width=5em, text centered, rounded corners, minimum height=4em]
\tikzstyle{line} = [draw, very thick, color=black, -latex']
\tikzstyle{cloud} = [draw, ellipse,fill=white, node distance=2.5cm,
minimum height=2em]

\begin{document}
	
	\title{Where does active travel fit within local community narratives of mobility space and place?}
	
	\author{
		\name{Alec Biehl\textsuperscript{a}, Ying Chen\textsuperscript{a,b}, Karla Sanabria-V\'{e}az\textsuperscript{c}, David Uttal\textsuperscript{d}, \& Amanda Stathopoulos\textsuperscript{a}} \thanks{Corresponding author: Amanda Stathopoulos. Email: a-stathopoulos@u.northwestern.edu}
		\affil{\textsuperscript{a}Department of Civil and Environmental Engineering, Northwestern University; \textsuperscript{b}Northwestern University Transportation Center; \textsuperscript{c}Department of Education, University of Puerto Rico, R\'io Piedras; \textsuperscript{d}Department of Psychology and School of Education and Social Policy, Northwestern University}
	}
	
	\maketitle
	
	\begin{abstract}
		 Encouraging sustainable mobility patterns is at the forefront of policymaking at all scales of governance as the collective consciousness surrounding climate change continues to expand. Not every community, however, possesses the necessary economic or socio-cultural capital to encourage modal shifts away from private motorized vehicles towards active modes. The current literature on `soft' policy emphasizes the importance of tailoring behavior change campaigns to individual or geographic context. Yet, there is a lack of insight and appropriate tools to promote active mobility and overcome transport disadvantage from the local community perspective. The current study investigates the promotion of walking and cycling adoption using a series of focus groups with local residents in two geographic communities, namely Chicago's (1) Humboldt Park neighborhood and (2) suburb of Evanston. The research approach combines traditional qualitative discourse analysis with quantitative text-mining tools, namely topic modeling and sentiment analysis. The aim of the analysis is to uncover the local mobility culture, embedded norms and values associated with acceptance of active travel modes in different communities. The analysis uncovers that underserved populations within diverse communities view active mobility simultaneously as a necessity and as a symbol of privilege that is sometimes at odds with the local culture. Thereby, this research expands on the walking and cycling literature by providing novel insights regarding the perceived benefits of, and barriers to, equitable promotion of these modes. The mixed methods approach to analyzing community member discourses is translated into policy findings that are either tailored to local context or broadly applicable to curbing automobile dominance. Overall, residents of both Humboldt Park and Evanston envision a society in which multimodalism replaces car-centrism, but differences in the local physical and social environments would and should influence the manner in which overarching policy objectives are met.
	\end{abstract}
	

\section{Introduction}
Community impact assessment is an integral component of transportation planning and policymaking. A 2018 report sponsored by the U.S. Federal Highway Administration purports that ``the assessment of community impacts supports sustainable, livable communities; promotes community values and thriving neighborhoods; and contributes to economic growth and general well-being'' \citep{Grant2018}. Yet, notable gaps persist in the literature. There is a need for more research on the interplay among contextual factors, power dynamics, and collaboration between government actors at the level of local neighborhood residents in determining the feasibility of innovative transportation projects \citep{Marsden2017}. There is growing awareness of the need to collect data and compare active mobility behaviour at the city-level \citep{gerike2016physical}. In addition to physical place, active mobility culture is also generated as a social and flexible space \citep{aldred2012constructing}. These two dimensions meet at the local community level  where citizens navigate active mobility decisions. Mobility scholars therefore need to improve analytical frameworks and data collection methods so that research participants, along with values and norms related to the communities they belong to, are better represented. Taking a local community perspective promises to improve transport policy design and implementation (see, for instance, Machler and Golub (\citeyear{Machler2012})). Accounting for socioeconomic diversity and the needs of traditionally disadvantaged groups are of particular importance, as much research relies on responses from individuals who are white, middle- to high-income, and/or college-educated \citep{Cavoli2015}. \par

A promising way forward is to use qualitative research methods to construct a critical dialogue among community members regarding potential paths towards low-carbon societies, particularly with respect to walking and cycling behaviors \citep{Ferrer2015, Segar2017}. At an aggregate level, the proliferation of active travel modes has the potential to help foster greater levels of physical activity, social interaction, and overall `healthier' lifestyles, for example with respect to nutritional intake \citep{Dia2017}. These outcomes are conditional on a supportive built environment, whose relationship with travel behavior is well-studied \citep{Wang2017, Mokhtarian2008, Handy2005}. Thus, there is sufficient evidence in the literature indicating the critical influence of both land use patterns and infrastructure availability on human movement, especially regarding active transportation. Pedestrians require close proximity to a variety of destinations and opportunities to fulfil their daily needs \citep{McDonald2012, Wood2010}. Cyclists, while less restricted by distance and speed, must be permitted a travel environment that assuages safety concerns against car traffic, accomplished through the installation of protected bike lanes or traffic calming measures \citep{Motoaki2015, Heinen2010}. The causal mechanisms behind modal shifts via physical `hard' policy measures remain obfuscated, though; it is still unclear whether individuals choose to live in neighborhoods that support their travel preferences or if interventions in the built environment actually motivate behavioral change \citep{Schwanen2005}.

While early research on active mobility used simpler representations, more recent work integrates travel behavior theories more thoroughly \citep{gotschi2017towards}. This paves the way for designing `soft' policies that are grounded in psychological theory, ideally implemented in tandem with traditional approaches to encourage the adoption of sustainable mobilities \citep{Bamberg2011}. In addition, the notions of \textit{subjective well-being} (i.e. self-reported happiness and satisfaction with life domains) and \textit{social norms} (i.e. behavioral expectations within a society or group) are becoming more prominent in the transportation literature in response to growing dissatisfaction with the normative assumptions of expected utility theory. Fundamental to behavioral economics \citep{Metcalfe2012} and sociology \citep{Cairns2014}, there is growing agreement that (a) the primary goal of transport policy should be the improvement of individual and community well-being, and (b) travel behavior change campaigns have a greater chance of being successful if policymakers account for the norms and values embedded in the local social environment. Defining and encapsulating these concepts with sensitivity to context, however, remains a challenge for researchers and practitioners seeking to curb private vehicle use. \par

Conducting focus groups could illuminate the community specificities required for effective policy design. A focus group serves as either a standalone source of information \citep{Nielsen2015, Lo2013} or as part of a mixed-methods design, in which the qualitative analysis informs a survey instrument or complements quantitative data analysis \citep{Ruiz2014}. It is pertinent that more travel behavior researchers recognize the value of incorporating qualitative methods to understand motivations and complexities underpinning mobility choices \citep{Grosvenor2000,clifton2003qualitative}. Recent years have seen an increase in the use of focus group approaches in the transportation field, to investigate such phenomena as car and public transit perceptions \citep{Hagman2003,Beirao2007,Guiver2007}, walking behavior \citep{Segar2017,Ferrer2015}, transport disadvantage \citep{Combs2016}, vulnerable groups \citep{Lubin2016,Race2017,Adorno2016}, ride-sharing perceptions \citep{Nielsen2015}, use of two-wheelers \citep{Hagen2016}, and the importance of accounting for local context \citep{Biggar2017a}. \par

Hence our research utilizes focus group discourses to glean insight into how community-specific factors shape norms and beliefs about (active) mobility innovation. Our research advances the field in two main directions, providing new theoretical insights, and using mixed qualitative and quantitative analysis. \textit{First}, we explore the role of the socio-cultural community context for the adoption of walking and cycling by different communities. We collect focus group dialogue from two distinct community areas: \textit{Humboldt Park}, Chicago's historically Puerto Rican neighborhood located approximately five miles west of downtown, and \textit{Evanston}, a majority-white suburb along Chicago's northern border that possesses a mixture of urban and suburban features. The focus group design aligns with the goals delineated in the literature investigating travel behavior interventions \citep{Davies2012, Bamberg2011, Scheepers2014}. This is important given that neighborhood-based research has shown that ethnic groups tend to embody distinct mobility cultures \citep{Klinger2013, Lin2008, Rietveld2004}. Specifically, our research contributes to understanding diverse local context factors and translating them into policy recommendations that align with community goals.

\textit{Second}, we use a mixed methods approach to analyze the data, relying on highly qualitative discourse texts derived from the focus groups, merged with text-mining approaches that have not been used in any previous mobility focus group analysis. Specifically, we employ topic modeling and sentiment analysis algorithms to find patterns in the discourse data. The approach ensures deep contextual understanding, unified with formal analysis tools to ensure less arbitrariness of interpretation. This approach is also uniquely suited to compare multiple areas in line with the need to explore different types of communities unique acceptance and cultural values related to active mobility. Joining the carefully executed community focus group consultations with text-mining approaches paves the way to identify an `optimal' level of policy tailoring across geographic scales, i.e. a balance between ubiquitous and area-specific strategies.  \par

The paper is structured as follows. The next section describes our study design and methods, including the recruitment process and guidelines for conducting the focus groups; then, we provides brief historical and socio-demographic details of Humboldt Park and Evanston. Subsequently, we elucidate the extracted topics, which empower traditional content analytical interpretations of the transcriptions, followed by a presentation of sentiment scores for different travel modes, which serve to represent key affective features of the group dialogue. The final two sections discuss the implications of our findings and provide insights for transport policy that balances generality and specificity.

\section{Research Methods}
This research contributes to the current state of knowledge by examining the perceptions and values tied to community area mobility, for which semi-structured focus group discussions are an ideal source of data. We propose a novel method to analyze this data. To expound, unlike existing transportation focus group work relying on qualitative grounded theory or discourse analysis---typically presenting results in the form of quotes followed by interpretation---we apply text mining tools to generate results for critical reflection on the data set. By introducing these tools as part of the analytical repertoire, we hope to offer a better basis of comparison across community area contexts, as well as streamlining (and rendering more transparent) the investigative practices for `personalized' policy innovation.

\subsection{Setting}
We conducted a total of five focus group sessions in two communities. Three focus groups took place in Humboldt Park ($n$ = 15) over the course of two weeks in July 2016. Two of them were held at the Puerto Rican Cultural Center, a grassroots organization that serves to preserve local cultural identity, while the Humboldt Park Diabetes Empowerment Center hosted the third focus group. In Evanston, two focus groups ($n$ = 9) took place during one week in September 2016, with recruitment assistance from the city's Transportation and Mobility coordinator. The first focus group took place at the Evanston Civic Center while the second was held in Northwestern University's Chambers Hall. These locations were primarily chosen for their accessibility to focus group participants.

\subsection{Participants}
To assuage privacy concerns, we did not collect specific identifying information on focus group participants, but make the following observations. Humboldt Park participants either held a leadership role in the Puerto Rican Cultural Center, or were an employee of a business or organization located along the Paseo Boricua, the heart of the local Puerto Rican community. All individuals were of Hispanic origin and collectively represented the entire 18+ age spectrum. Meanwhile, the Evanston focus groups consisted of bike shop owners, local professionals, and members of government-sponsored task forces interested in transportation sustainability and accessibility issues; participants were middle-aged or elderly white individuals. The representation of older citizens and residents of deprived community areas has been identified as a priority area in research linking mobility and well-being \citep{Cavoli2015}. Hence, our representation promises to advance less understood perspectives. Moreover, our analysis did suggest a large diversity of participants knowledge and opinions regarding the discussion topics. Thus, we believe that, despite possible issues over sample representativeness, the participant testimonials offer useful insights about individual- and neighborhood-level perceptions of (active) mobility innovation. \par

\subsection{Study Design and Procedures}
The open-ended nature of qualitative data collection allows researchers to gain intricate details to help identify the factors relevant for active transportation use and acceptance in local community areas. Compared to formal surveys and interviews, whose direction is notably dependent on the researcher's predetermined intellectual inquiry, the focus group permits a more organic relaying of information through participant interactions with the researcher as discussion facilitator. Focus group research is more effective for garnering contextual knowledge, particularly with respect to unanticipated information arising from the ensuing dialogue \citep{hughes2002}. \par

We obtained written consent, in accordance with local IRB guidelines, regarding individuals' participation, the audio recording of the discussion, and the use of anonymous direct quotes or paraphrased material in any potential research publications. To establish an amicable environment for the focus groups, the moderators facilitated brief introductions before delving into the procedures and expectations for the ensuing discussion. In the case of Humboldt Park, we also encouraged participants to speak in Spanish if they had trouble expressing their thoughts and opinions in English, since one of the moderators was bilingual and translated this dialogue for the written transcriptions. \par

Each focus group lasted approximately 90 minutes and had two members of the research team present to act as moderators: one person was in charge of facilitating the discussion while the other person took supplementary notes to complement the audio recordings. A mapping exercise was used as an ice-breaking activity, where participants marked out local transportation infrastructure and briefly discussed observed differences. This set the foundation for discussion of how different transportation modes impact life in the local community area. Ten questions, stemming from four general topics, comprised the focus group discussion guidelines: [T1] perceptions of the built environment; [T2] physical and psychological well-being issues; [T3] cultural and community identities; and [T4] ideas to improve local mobility services. The full discussion guide is presented in the Appendix. Also, we note that the appearance of HP/EV denotes the specific mention of Humboldt Park/Evanston. In line with the fluid nature of the study design and time constraints, a sub-question could be skipped if the moderators judged that a sufficient amount of time was spent on one of the four major topics. The discussion guide also had built-in flexibility so that, if a participant began addressing a question the moderators had not yet posed, they could capitalize on the moment to move the discussion in a new direction.

\subsection{Data Processing}
The focus group audio recordings were transcribed by a professional company, with the exception of Spanish dialogue that the third author of this paper transcribed. After all transcriptions were finished, the first author fixed spelling errors and removed filler words. The prepare for analysis, copies of the transcriptions were edited in Microsoft Excel removing words spoken by the moderators so that the applied machine learning algorithms only utilized the contributions of participants. The original transcripts was used to understand what simple yes/no responses corresponded to, thus enhancing the text mining results with our own insights.

\section{Overview of Focus Group Research Sites}
\begin{figure}[h!]
	\centering
	\includegraphics{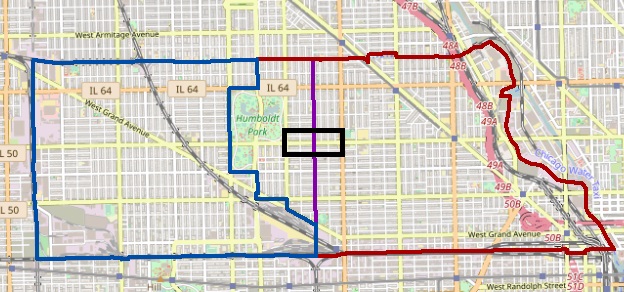}
	\caption{Several places in Chicago are associated with the name \textit{Humboldt Park}. Outlined in blue is the officially designated community area of Humboldt Park. If, however, we extend the eastern border of the Humboldt Park community area to the purple line, we obtain the historic Humboldt Park neighborhood, which overlaps with the West Town community area, shown with a red border. The famous \textit{Paseo Boricua}, a one-mile stretch of Division Street that is the historic epicenter of Puerto Rican culture and indicated by the black box, thus straddles geographic ambiguity. Finally, the public park after which the community area and neighborhood are named is shown in green. Created using ArcMap Version 10.5.}
	\label{fig:HP}
\end{figure}

\begin{figure}[h!]
	\centering
	\includegraphics[scale = .67]{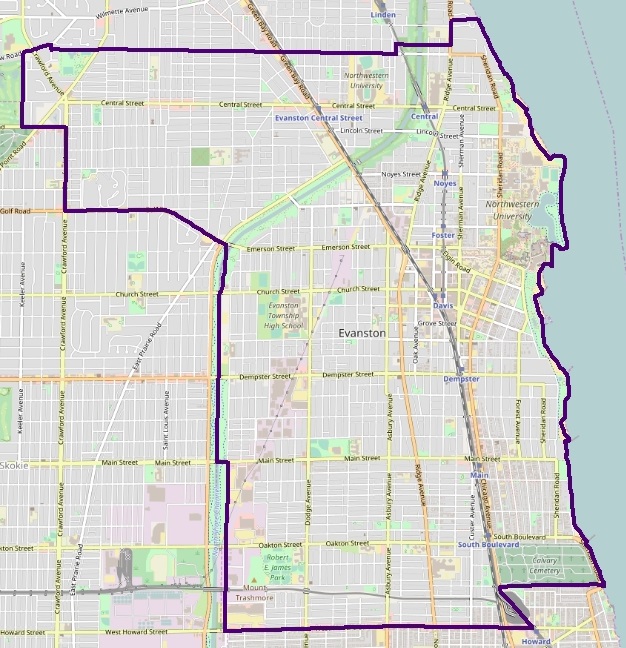}
	\caption{The city of Evanston is shown within the purple border. It is home to Northwestern University and is served by the CTA `L' Purple Line (rail transit) and Metra Union Pacific North Line (commuter rail), which are visible in the map. Created using ArcMap Version 10.5.}
	\label{fig:EV}
\end{figure}

By selecting community areas that are fundamentally different from one another in terms of demographic and socioeconomic composition, we have the opportunity to highlight points of convergence and divergence that would impact transport policy at different scales of implementation. Therefore, before delving into data analysis, it is important that readers are familiarized with Humboldt Park (Figure \ref{fig:HP}) and Evanston (Figure \ref{fig:EV}) in order to understand the incarnations of space and place with which the focus group participants likely interact. Table 1 provides sociodemographic information associated with the community areas of interest from documents produced by the Chicago Metropolitan Agency for Planning in June 2016. Compared to Evanston, Humboldt Park has a higher percentage of Black, Hispanic, and unemployed residents, versus a lower percentage of college-educated residents and walking/cycling commuters. Additionally, the median household income is less than half that in Evanston. \par

The contrasting characteristic sets of these two community areas are potentially suggestive of different interpretations regarding the role of active travel modes in their respective futures. On the other hand, shared perspectives amongst focus group participants would highlight consensus on certain factors. \par

\begin{center}
	\begin{threeparttable}
		\scriptsize
		\captionsetup{justification=centering}
		\caption[labelsep=period]{American Community Survey Statistics for Community Areas}
		\begin{tabular}{|>{\centering\arraybackslash}m{4cm}||>{\centering\arraybackslash}m{1.5cm}|>{\centering\arraybackslash}m{2.5cm}|@{}m{0pt}@{}} 
			\hline
			\textbf{Variable} & \textbf{Evanston}\tnote{1} & \textbf{Humboldt Park}\tnote{2} &\\ [5pt]
			\hline \hline
			Total Population & 75,282 & 54,515 &\\ [5pt]
			\hline
			\% Black & 17.2 & 41.5 &\\ [5pt]
			\hline
			\% Hispanic & 10.1 & 51.5 &\\ [5pt]
			\hline
			\% Age 20-34 & 24.3 & 24.6 &\\ [5pt]
			\hline
			\% Age 35-49 & 19.5 & 18.6 &\\ [5pt]
			\hline
			\% Age 50-64 & 17.9 & 15.1 &\\ [5pt]
			\hline
			\% Age 65-79 & 8.2 & 5.7  &\\ [5pt]
			\hline
			Median HH Income & \$69,347 & \$32,484 &\\ [5pt]
			\hline
			\% Bachelor's degree or higher & 66.4 & 12.7 &\\ [5pt]
			\hline
			\% Occupied HH & 91.1 & 83.4 &\\ [5pt]
			\hline	
			\% Unemployed & 8.2 & 17.5 &\\ [5pt]
			\hline
			\% Transit Commuter & 23.5 & 24.1 &\\ [5pt]
			\hline
			\% Walk or Bike Commuter & 17.4 & 5.7 &\\ [5pt]
			\hline			
		\end{tabular}
		\begin{tablenotes}
			\item[1] http://www.cmap.illinois.gov/documents/10180/102881/Evanston.pdf
			\item[2] http://www.cmap.illinois.gov/documents/10180/126764/Humboldt+Park.pdf
		\end{tablenotes}
	\end{threeparttable}
\end{center}

\subsection{Humboldt Park}
The success of community-based research relies on strong connections with the people that live and work there. Accordingly, the selection of Humboldt Park as one of the focus group sites was made possible by the fact that two of the authors performed research in the neighborhood during the previous year and maintained connections with local leaders and residents. \par

Puerto Ricans are the second largest Latino ethnic group in the United States, which from 1970 to 2010 contributed a total of 4.7 million people to the country's population  \citep{Cintron2012}. This mass migration has an important socioeconomic and historical background due to Puerto Rico's colonial status with the United States. Juffer (\citeyear{juffer2006}) states that ``Puerto Ricans were basically resettled in Chicago by the U.S. government in the 1940s and '50s, recruited to work in the steel mills and factories during the boom economy of World War II'' (p. 146). By the '60s Puerto Ricans were highly concentrated in Lincoln Park, West Town, and Humboldt Park neighborhoods \citep{PuertoRicans}. As of 2010, Chicago had 196,913 Puerto Rican residents, of which 19\% lived in poverty and 27\% of households received food stamps, compared with 8\% of white families in Chicago \citep{Cintron2012}. These statistics portray how Puerto Rican residents face barriers, like many ethnic minority groups in the United States, in achieving upward socioeconomic mobility. 

Humboldt Park has been the focus of a handful of local initiatives that aim to enhance health education and promote active lifestyles as part of recognized needs to improve the overall quality of life \citep{HPQoL} and offer better access to digital infrastructure \citep{HPSCP}. To expand, in a report by Northwestern University's Feinberg School of Medicine \citep{ChicagoHealth}, major health concerns in the community area include childhood obesity, the spread of HIV via injection drug use, high rates of diabetes, and more frequent breast cancer screening. It is plausible that active transportation could function as a quintessential resource to improve physical, and even mental, health of residents, but city planners must remain cognizant of the local ethnography to be welcomed and respected as change agents. \par

Humboldt Park has a strong history of resistance against gentrification in order to preserve Puerto Rican heritage and that of other marginalized groups comprising the community area's social capital. A key player within this narrative is the Puerto Rican Cultural Center, which has been persistently vocal in protecting the physical and cultural manifestations that represent and influence individual and neighborhood identities \citep{Rinaldo2002}. The connection between space and place in Humboldt Park is expressed through numerous artistic renderings of Puerto Rico's historical struggle with colonialism, to such a degree that neighborhood tours evolved into a means of connecting with privileged populations while affirming ethnic identity. \par

Humboldt Park thus stands out as an opportunity for transportation researchers to understand the contextual forces that impact acceptance of policy and mobility innovation \citep{Santos2007}. Recently, Lubitow, Zinschlag, and Rochester \citeyearpar{Lubitow2015} demonstrated through a series of interviews in Humboldt Park that the idea of expanded bicycle infrastructure is not opposed in the community area. Yet, its association with so-called \textit{hipsters} and \textit{yuppies}, as well as white males, is a hindrance to cycling growing as a form of daily mobility. We refer readers to Wilson and Grammenos \citeyearpar{Wilson2005}, for more information regarding Humboldt Park and discourse surrounding gentrification.

\subsection{Evanston}
Evanston is the gateway to Chicago's North Shore suburbs, but offers a sense of urban living through its `smart growth' approaches to land use and infrastructure development. According to its Transportation and Mobility website, the city is committed to 1) further developing its active transportation network, 2) integrating this with its transit and on-demand transportation services, and 3) tackling the first- and last-mile access/egress issue through the expansion of Divvy bikeshare \citep{TransportMobility}, which consists of 11 local stations as of 2018. Thus, it is evident that Evanston continues to strive to be a leader in multimodalism, and two recent accomplishments by the city are worth noting. \par

First, Evanston is one of eight recipients of the Round Four Transit Planning 4 All grant, a joint effort by the Federal Transit Administration and the Administration for Community Living to support investment in transport initiatives that address the lack of accessibility for the aging and physically-disabled populations. The focus of the project is to determine appropriate solutions for the first- and last-mile travel needs of the target populations. Project partners include PACE Bus, which also offers paratransit, and organizations whose missions are to improve the quality of life for elderly and disabled individuals \citep{TP4A}. Second, Evanston is a recognized leader in sustainability from the STAR Communities program, receiving the prestigious full rating that only Tacoma, Washington and Broward Country, Florida also possess. The purpose of this program is to foster initiatives in participating community areas that encourage goals that improve livability in the following categories: the built environment, the natural environment, climate considerations, equity and empowerment, health and safety, arts and education, and the economy \citep{STARComm}. This holistic perspective on sustainability aligns with the major topics guiding our focus group discussions.

\section{Topic Modeling}

\subsection{Theory}
As data sets continue to grow in size and complexity, it becomes increasingly challenging to find and discover useful information for answering pressing research questions. In response, new analytical tools have been developed to help researchers organize, search, and understand large datasets. One example is topic modeling, a widely used data mining technique for discovering the hidden semantic structures in textual information, such as a collection of related documents \citep{steyvers2007probabilistic}.\par
In the transportation analysis setting, Sun and Yin (\citeyear{sun2017discovering}) use topic modeling to analyze articles published in transportation journals over a 25-year period to reveal the regional patterns of research sub-fields/topics. Similarly, Gatti et al. (\citeyear{gatti2015historical}) develop a topic model to reveal the temporal evolution of themes in the operations research and management science fields.

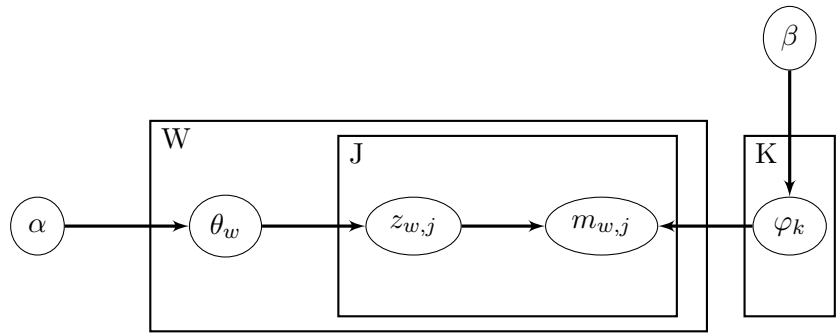
\begin{figure}[h]
\centering
\begin{tikzpicture}[scale=2, node distance = 2cm, auto]
\node [cloud] (one) {$\alpha$};
\node [cloud, right of=one] (two) {$\theta_{w}$};
\node [cloud, right of=two] (three) {$z_{w,j}$};
\node [cloud, right of=three] (four) {$m_{w,j}$};
\node [cloud, right of=four] (five) {$\varphi_{k}$};
\node [cloud, above of=five] (six) {$\beta$};

\path [line] (one) -- (two);
\path [line] (two) -- (three);
\path [line] (three) -- (four);
\path [line] (five) -- (four);
\path [line] (six) -- (five);

\draw[thick] ($(five)+(-0.3,0.6)$) rectangle ($(five)+(0.3,-0.6)$) node[pos=0, yshift=-5mm] {K};
\draw[thick] ($(three)+(-0.5,0.6)$) rectangle ($(four)+(0.5,-0.6)$) node[pos=0, yshift=-5mm] {J};
\draw[thick] ($(two)+(-0.5,0.7)$) rectangle ($(four)+(0.7,-0.7)$) node[pos=0, yshift=-5mm] {W};
\end{tikzpicture}
\caption{Graphical depiction of LDA topic modeling algorithm.} \label{fig:LDA}
\end{figure}

The algorithm used in this paper is called Latent Dirichlet Allocation (LDA), shown in Figure \ref{fig:LDA}, and is a generative probabilistic model for collections of documents \citep{blei2003latent}. The basic idea is that \textit{W} documents are represented as random mixtures over \textit{K} latent topics, where each topic is characterized by a distribution of words. For each document \textit{w} with length \textit{$N_{w}$} in a text corpus \textit{D}, the steps of LDA are as follows:

\begin{enumerate}
	\item Determine the topic distribution of each document \textit{w}: choose $\theta_{w}$ $\sim$ Dir($\alpha$), \textit{w} $\in$ $\lbrace$1,2,...,\textit{W}$\rbrace$. Dir($\alpha$) is a Dirichlet distribution with symmetry parameter $\alpha$, which is typically sparse (i.e. less than 1).
	\item Determine the word distribution of each topic \textit{k}: choose $\varphi_{k}$ $\sim$ Dir($\beta$), \textit{k} $\in$ $\lbrace$1,2,...,\textit{K}$\rbrace$. Dir($\beta$) is a Dirichlet distribution with symmetry parameter $\beta$.
	\item For each of the word positions (\textit{w},\textit{j}), where \textit{j} $\in$ $\lbrace$1,2,...,$N_{w}$$\rbrace$:
	\begin{enumerate}
		\item Choose a topic \textit{$z_{w,j}$} $\sim$ $Multinomial_{K}$($\theta_{w}$).
		\item Choose a word \textit{$m_{w,j}$} $\sim$ $Multinomial_{V}$($\varphi_{\textit{w},\textit{j}}$).
	\end{enumerate}
\end{enumerate}

To clarify, this algorithm assumes a fixed set of topics comprising the documents under investigation, governed by a specified Dirichlet mixture distribution. Each document is `generated' by a sequential process in which a topic followed by a word are outcomes of draws from separate multinomial distributions. This essentially replaces the `qualitative factor analysis' process described in Nielsen et al. (\citeyear{Nielsen2015}), employed by the authors to identity themes within the focus group discussions.

\subsection{Application: Theme Identification}
We applied the algorithm presented in the previous subsection to each community area's set of transcriptions. Table 2 shows the results of this analysis. To ensure meaningful output, we removed `things' and `stuff' from the inquiry in addition to combining words with the same base (e.g. `walk' and `walking') and singular/plural cases. Notice that the algorithm repeats some words across topics; this is not problematic, though, because the use of these words are different throughout the dialogue. Based on this output, the research team searched the transcriptions for (groups of) sentences containing one or more of the words to identify the text subset corresponding to the implied theme, which is propounded using researcher intuition regarding the conversational contents. Thus, topic modeling is used as a component of, rather than a substitute for, traditional content analysis in the qualitative literature. Each of the following ten subsections synthesize the discussion among focus group participants with respect to the identified topics and themes. \par
\vspace*{-\baselineskip}
\begin{center}
	\begin{threeparttable}
		\scriptsize
		\captionsetup{justification=centering}
		\caption[labelsep=period]{The five topics as represented by extracted word groupings for each community area.}
		\begin{tabular}{|>{\centering\arraybackslash}m{2cm}||>{\centering\arraybackslash}m{5.55cm}|>{\centering\arraybackslash}m{5.5cm}|@{}m{0pt}@{}} 
			\hline
			\textbf{Topic Number} & \textbf{Evanston (EV)} & \textbf{Humboldt Park (HP)} &\\ [5pt]
			\hline \hline
			\ Topic 1 & bus, walk, good, kind, beach, lake, back & Puerto Rico, transportation, public, Humboldt Park, walk, make, space &\\ [5pt]
			\hline
			\ Topic 2 & street, bike, time, east, center, parking, community, make & feel, work, back, home, car, bus, place, safe, Western &\\ [5pt]
			\hline
			\ Topic 3 & a lot, school, car, kids, place, traffic, work, issue, Target, side & a lot, culture, years, good, live, bike, Uber, idea &\\ [5pt]
			\hline
			\ Topic 4 & Evanston, people, transportation, downtown, grocery, great, area, neighborhood & bike, ride, city, Chicago, money, time, neighborhood &\\ [5pt]
			\hline
			\ Topic 5 & west, Chicago, car, live, stop, city, north, bus, bit & people, community, neighborhood, bike, bus, part, lanes &\\ [5pt]
			\hline
		\end{tabular}
	\end{threeparttable}
\end{center}

\subsubsection{HP Topic 1: Space-making and evaluation compared to Puerto Rico.}
The residents of Humboldt Park are fierce defenders of their neighborhood space, defined by numerous symbols and artistic renderings of Puerto Rican heritage. These typically serve as reminders of home for migrants and as learning opportunities for youth to embrace their cultural heritage. Resulting from this pronounced sense of autonomy is a keen awareness of the burgeoning conflict over ``shared space'' as it relates to different mobility cultures \citep{Klinger2015}. In the warmer months, many restaurants and caf\'{e}s will claim part of the sidewalk for customer use, which infringes on pedestrian space. This, according to one participant, is viewed by some as the privatization of a public good. Additionally, in the absence of dedicated bike lanes, this increases the tension between pedestrians and cyclists who do not feel comfortable using streets. As the presence of cyclists grows larger, even when bike lanes exist, there tends to be resistance from car users who are forced to be more aware of their environments and must change habitual driving behaviors. Therefore, the multimodal setting, coupled with a stubbornness to adapt to a changing built environment, engenders conflict over the rights of accessibility to shared spaces. \par

There are, however, points of praise regarding transportation services in Humboldt Park, including comparative benefits to what is available in Puerto Rico. Many participants are content with public transportation, as bus service is frequent and the network is easy to navigate. Furthermore, the nearest rail transit station is `accessible' by roughly 30 minutes of walking. Although medical facilities are plenty, Humboldt Park suffers from its status as a food desert and residents must travel to adjacent neighborhoods for fresh produce. The Humboldt Park green space is a point of attraction for both locals and visitors alike, offering a ``critical space of wellness'' for those seeking an escape from the stresses of urban life. Overall, the travel experience in Humboldt Park is more comfortable than in Puerto Rico, mainly due to more efficient operations and perceptions of personal safety.

\subsubsection{HP Topic 2: Barriers to active transportation use and fulfillment of well-being.}
While there was general agreement that an increased presence of walking and cycling is good for society, respondents gave considerable attention to concerns about their own participation in the adoption of these modes. The primary barrier is fear for personal safety. Several comments highlighted the dangers intersections pose because drivers are either unaware of or aggressive toward cyclists. Bike lanes are typically located next to parking spaces, which raises the potential issue of getting hit with car doors as people exit, and the network is not contiguous, which presents challenges for navigating the cityscape. This set of issues is particularly alarming for families: the law forbids children under the age of 12 from riding a bicycle in the street, but sidewalks with high levels of pedestrian traffic are not viewed as much safer. For women, the threat of harassment is a major deterrent of walking activity, so cycling is a more appealing option owing to the additional speed it offers. \par

Other barriers prevalent in Humboldt Park relate to the accumulation of neighborhood resources designed to improve individuals' life satisfaction. While a strong sense of place attachment and identity unites local residents, and is recognized as critical for good mental health, an endemic lack of individual economic prosperity simultaneously detracts from efforts to enhance neighborhood-level well-being. To demonstrate, several respondents use the neighboring community areas of Logan Square and Wicker Park, which are booming with development yet highly gentrified, as examples of what Humboldt Park is \textit{not}. Within the focus groups, this discussion is placed in the broader debate of individual vs. collective interest, with the latter mindset designated as a strong value in Latino culture. Thus, returning to the topic of transportation, active travel modes are viewed mostly as conduits of leisure fulfillment, which improves individual well-being, while improvements to public transportation are viewed as a top priority for meeting utilitarian needs of the community area. This is because the majority of jobs and daily amenities are located outside of the neighborhood.

\subsubsection{HP Topic 3: Multi-faceted cultural presence and reaction to innovation.}
The notion of \textit{culture} contains complex nuances in the social environment of Humboldt Park, a topic that is not well-studied in the travel behavior literature \citep{Wang2017, Rietveld2004}. Beyond the physical embodiments of ethnic pride, namely the murals and 60-foot Puerto Rican flags signaling entry into the \textit{Paseo Boricua}, the focus groups touched on several cultural aspects that define neighborhood lifestyles and mobilities. One focus group respondent posed the idea of initiating neighborhood walking tours of the aesthetics aspects of the built environment, mainly to forge a `sense of community' between locals and non-locals. While the neighborhood is highly walkable, there is a limited yet growing cycling presence in Humboldt Park, mainly due to its absence in Latino households as a utilitarian activity. Concurrently, there are a couple of well-known cycling clubs, such as the \emph{Chicago Cruisers}, that are demonstrating the social atmosphere of active travel engagement. Moreover, a few bicycle shops, such as \emph{Ciclo Urbano}, are working with youth to build skill sets relevant to the job market. Tied to this subculture is an emerging awareness over public health issues: the neighborhood is home to several centers whose purpose is to improve the physical well-being of locals through educational programming. However, a conscious merging of active travel and public health remains elusive, which Cavoli et al. (\citeyear{Cavoli2015}) note is difficult to assess in data collection where the researcher is not present to offer clarification. \par

What is important to recognize from the focus group dialogue is the dichotomy between internally-motivated efforts and acceptance of innovation from outside the community area. A common example is the growing presence of ride-sourcing companies such as Uber and Lyft. Historically, taxi service in Humboldt Park has been sparse and expensive in terms of the destinations associated with its use, mainly the airports and places difficult to reach by public transportation. Thus, even though ride-sourcing presents a cheaper alternative for the same trips, there are two major roadblocks to its widespread adoption in Humboldt Park. First, a couple of participants commented on the longer wait times for pick-ups compared to other Chicago neighborhoods, due to limited driver interest in serving the community area. Second, relatively few individuals in Humboldt Park have the resources (i.e. mobile device, credit cards) or awareness to participate in this mobility service, and one participant commented on the need to invest in social marketing campaigns tailored to local interests.

\subsubsection{HP Topic 4: Humboldt Park within the tapestry of Chicago neighborhoods.}
The neighborhood composition of Chicago is distinct, which is partially due to its history of segregation based on socioeconomic status. As one respondent puts it, \textit{``Chicago is the most segregated city... whether people see it as good or bad because you do go into a certain neighborhood and you get the culture, you get the feel of it. However, it is so concentrated with just that culture that other people might feel overwhelmed to even visit it. It's so segregated. Again, it could be a good thing or a bad thing, but I think that's what makes certain neighborhoods strong, or stronger.''} This idea is reiterated in the context of gentrification, a phenomenon that Humboldt Park has resisted for decades to preserve place identity; this could easily evolve into resistance to travel behavior change if it is perceived as connected to this oppressive socio-historical phenomenon \citep{Davies2012}. Two important points emerge from this discussion. First, while socioeconomic integration is vital to the pursuit of equality, there is disagreement as to what the ideal scenario would be. On one hand, mixed-income development could diversify a neighborhood, but many contemporary investments appear to gentrify minority neighborhoods, as market protections for these households usually do not exist. On the other hand, to protect the cultural integrity of neighborhoods, some participants like the idea of distinct places that are interconnected through a well-functioning transportation network that allows non-residents to easily access what various community areas have to offer. Second, the interplay between transportation and gentrification is contentious. Some respondents believe in total disassociation of the two topics, as the issue lies more with communication of information to disadvantaged groups, while others view the expansion of transportation services as potential conduits of gentrification, since the flow of movement is viewed as lopsided. In other words, mobility innovation could be associated with catering to the desires of non-residents wanting greater access to the neighborhood rather than providing better access to amenities outside the neighborhood for current residents.

\subsubsection{HP Topic 5: Active participation in community preservation and progress.}
Due to prevailing displeasure with the spatial dominance of vehicles, the overall sentiment is that development should focus on eco-friendly neighborhood revitalization that boosts morale through improvements to the design of train stations and bus stops, as well as the `social status' associated with active travel modes; these would promote traffic calming and collectivism. Involving local residents in the transportation planning and policy-making process is seen as a missing yet integral component for encouraging new forms of travel . For instance, in the case of Chicago's Divvy bikeshare, the idea is welcomed by community area leaders for its potential in encouraging healthier lifestyles, but its current usage is low because many residents feel that they did not `consent' to its presence in Humboldt Park and generally remain unaware of the enrollment process, including the discount for low-income users.

\subsubsection{EV Topic 1: Multimodalism with considerations for health and equity issues.}
All participants in the focus groups touched on Evanston's transition to a city that accommodates public and active modes of transportation. However, this process involves a `learning curve' for travelers to adjust to the emerging dynamics of the network. In particular, bus and pedestrian activity comprised much of the discussion surrounding the needs of the community area. Regarding the former mode, there was a general agreement that the focus of the public bus operations is to move people through the town rather than within it, which spurs a desire for an internal circulator system that better connects local amenities. However, a couple of participants noted that Evanston possesses a significant private bus fleet that serves specific destinations, such as hospitals or community centers, but information regarding these services has not been effectively disseminated. Furthermore, buses are subject to twofold stigmatization: they are `anti-social' and are associated with low social status. Meanwhile, despite the praise for increased walkability along the lakefront and downtown area that induces a sense of `being on vacation,' there is growing concern for the age-friendliness of active travel infrastructure, namely that poor maintenance (especially during the winter months) poses a threat to the physical health of senior citizens. \par

Concerns regarding well-being and equity also arise as participants discuss the need for people-focused transportation planning that is more concerned with social benefit over hard statistics. One individual highlighted the historic inequity in east-west transport services that deprives traditional disadvantaged neighborhoods from access to lakefront amenities. These `isolated subcommunities' are essentially excluded from conversations surrounding Evanston's growing multimodalism and the ability to improve one's subjective well-being by reducing car use. Another participant suggested that, in an ideal world, everyone should \textit{``get lost on a new bus route''} in order to appreciate its service to people from different walks of life.

\subsubsection{EV Topic 2: Mobility tensions and the geography of space and time.}
A natural result of movement away from a car-centric community area is emerging tension between users of different modes as street space is relearned and renegotiated among mobility cultures \citep{Klinger2015}. Cyclists feel a particular displacement as increased walkability induces more pedestrian activity on sidewalks and car traffic is often blind to or inconsiderate of their presence. Additionally, the desire for more frequent bus service is at conflict with the cycle-ability of Evanston because the space at the nexus of street and sidewalk cannot support both bus stops and protected bike lanes; in turn, both of these modes potentially threaten vehicle parking along the street, which concerns physically-disabled individuals who require accessible drop-off spaces as well as business owners who fear decline in business activity. Routine cyclists also point to the lack of a comprehensive bicycle network in Evanston, particularly along east-west corridors, and the notable anger from residents who live along streets with bicycle paths, such as Dodge Avenue. As these tensions continue to rise, sparking debate over the new rules of the road, the diffusion of innovative forms of mobility such as bikeshare remains uncertain. \par

Planners and policymakers must be cognizant of shrinking time budgets of individuals, as work requirements and burgeoning variety-seeking tendencies could discourage participation in slower travel modes compared to the private vehicle. For instance, Evanston has adopted a complete-streets policy that includes traffic dieting measures to improve the safety of active travelers, but participants claim these are often ineffective. In parallel, when older focus group participants reflect on past transit services available in Evanston, there is usually mention of a more relaxed pace of life during those times. Moreover, the propensity for individuals and policymakers to think/act in terms of short-term gains or losses deters the implementation and acceptability of changes that require long-term horizons to manifest and meet societal needs for the future, a sentiment echoed by Marsden and Reardon (\citeyear{Marsden2017}). One participant, who owns a bicycle shop in the city, points to dueling urban visions involving autonomous vehicles on one hand and active transportation on the other, expressing concern over how infrastructure and technological advancement will allow these modes to interact with one another in a manner preserving the safety of travelers.

\subsubsection{EV Topic 3: Concerns regarding car traffic and youth travel experience.}
The de-suburbanization of Evanston in favor of a more urban appeal is grounded in the narrative of a Target store located in the southwestern part of the city. While there is no debate surrounding its appeal, accessing the store is difficult without a car due to heavy traffic flow on Howard Street. One participant described the challenges of crossing the street to the eastbound bus stop to return home from her shopping trip, while another participant from a zero-car household expressed concern over riding a bicycle to the shopping plaza, especially with her children. Expanding on this point, there is some lament that many of today's youth are experiencing lifestyles that are heavily dependent on car use. This is particularly concerning in the context of after-school programs and activities that are difficult to access using public or active transportation, in addition to contemporary societal norms limiting the range of children's unsupervised travel compared to previous generations. Returning to the first EV topic, one participant hopes that schools and activity centers will eventually coordinate with one another to expose students to the notion one can be satisfied with travel and a lifestyle that is not heavily car-dependent.

\subsubsection{EV Topic 4: Reinvigorating a sense of community through the diffusion of social interaction opportunities.}
In contrast to the anti-social perceptions of public transportation described in the second EV topic, the increased walkability of Evanston is spurring greater interaction among community area members \citep{Wood2010}. Several participants alluded to the idea that the mere visibility of individuals going about their lives with the occasional small greeting is enough to bring a warm atmosphere to the city. However, multimodalism is only present in pockets of significant economic (downtown) or leisure (natural areas along the lakeshore) activity. Accordingly, these `hot spots' are somewhat disconnected from the rest of the city, both physically and mentally. To expand on the latter point, one result of the mapping activity was a discussion focused on the idea that residents living in southeast neighborhoods did not know of the amenities offered in northwest neighborhoods, and vice-versa. To improve intra-city socialization, publicly-sponsored events have recently been decentralized away from downtown, so that individuals have opportunities to mingle in new geographies and `fill in their local mental maps.' This discussion also involved brainstorming solutions that improve access to grocery stores for elderly individuals who cannot or do not use a car.

\subsubsection{EV Topic 5: Community pride through juxtaposition.}
Evanston residents tend to pride themselves on being inclusively progressive with roots in Midwestern conservatism, blending sustainability with socioeconomic diversity in a manner that surpasses comparable community areas within and near Chicago; for example, the Lincoln Park neighborhood, which is located just north of downtown. According to one participant, \textit{``We can be Lincoln Park north in terms of really great restaurants, really good shopping, really good transportation so that you can live here without having to necessarily have a car...[and] we don't have to have the bad parts of Lincoln Park, which is no parking.''} In response to this statement, another participant notes that there is very little diversity in Lincoln Park, due to a high cost of living, which is a much less severe problem in Evanston. Another comparison with the suburb of Oak Park illustrates that, while both cities are economically and ethnically diverse as well as situated at the end of the CTA rail transit network, the fact that Evanston is located adjacent to Lake Michigan is a deciding factor for attracting potential new residents. Nonetheless, as highlighted earlier, Evanston is still challenged with equity issues as it continues to embrace transit-oriented development and invest inclusively in new land-use patterns and infrastructure.

\subsection{Synthesis of similarities and differences}
In summary, the analysis of topics emerging from the focus group discussions reveals several similarities between Evanston and Humboldt Park, most notably the palpable place-based identities resonating in descriptions of the physical and social built environments \citep{Lengen2012}. Participants from both community areas emphasized the contentious nature of multimodalism through its requirement that travelers learn to negotiate a `shared space' inclusive of all modes. Emerging from this situation is a heightened awareness over safety concerns, especially regarding children \citep{Davies2012}. However, both community areas face dilemmas regarding the interaction of vehicle parking, bike lanes, and bus lanes at the nexus of the street and the sidewalk. Additionally, there is agreement that today's youth deserves a society that encourages eco-friendly mobility patterns, in terms of both activity accessibility and individual well-being. Participants also stressed that greater investment in public transportation should be a priority on a citywide scale. \par

There are, however, important differences to be noted. First and foremost, there is a clear disparity between Evanston and Humboldt Park regarding the economic, public health, and technological resources available to residents. Consequently, whereas participants from Evanston would typically point out what their community area \emph{possessed} compared to others, participants from Humboldt Park usually made comparisons based on what their community area \emph{lacked}. Another observation is that, although participants from both areas identified \emph{social status} as a barrier to travel behavior change, it was conveyed from distinct angles. In Evanston, the bus is considered to be the mode of low social status, which is in stark contrast to its utilitarian prominence in Humboldt Park. Meanwhile, Evanston residents tend to perceive utilitarian cyclists as \emph{looking down on others} while Humboldt Park residents tend to \emph{scoff at cycling} due to the mode's cultural irrelevancy beyond leisure fulfillment, which parallels the finding that low-income urban mothers tend to not view walking as physical activity \citep{Segar2017}. Finally, Humboldt Park focus group participants generally interpreted greater participation in active travel as a conduit of collectivism, while Evanston participants viewed this phenomenon as a liberation of constraints against individual lifestyle fulfillment. All of the above demonstrate the need to capture individual- and neighborhood-level variation in attitudes \citep{Heinen2010} and social norms \citep{Lois2015}, which are arguably among the most critical components of travel mode choice.

\section{Sentiment Analysis}
\subsection{Theory}
Another form of text mining is sentiment analysis, which embodies a set of computational methods to investigate and quantify the implied emotional facets of words and phrases. This type of analysis allows researchers to understand subjective information, notably opinions, more in-depth, thus providing a mechanism for which public policy and marketing strategy could adapt to the needs of individuals.

In order to quantify the positive, negative, and neutral sentiment degree in a given source text, the emotive phrases within a document (or paragraph) are extracted and scored; combining these scores produces the overall sentiment of the sentence. Importantly, the sentiment scoring process will score those sentences in the same way every time; since it is not affected by any other outside factors, this procedure is consistent. The scores are roughly between -1 and 1, but are often converted into a \textit{probability of positiveness}; that is, probabilities between 0.35 and 0.65 represent neutral sentiment, larger than 0.65 is positive sentiment, and less than 0.35 is negative sentiment. The procedure to score the sentiments in a document is described in the following steps:
\begin{enumerate}
	
	\item Building a document-term matrix (DTM), which is the result of vocabulary-based vectorization. The first step is to vectorize input documents by creating a map from words or n-grams to a vector space. An unsupervised algorithm called \textit{Paragraph Vector} \citep{le2014distributed} is adopted to generate vectors for sentence/paragraphs/documents in this study. Furthermore, vectorized texts have high efficiency in practice because they are stored as sparse matrices, thereby saving a lot of memory. \par
	
	\item Transforming the DTM. Term frequency-inverse document frequency (TF-IDF) transformation technique is applied to the DTM built in the first step. Since lengths of the documents/sentences can be different, TF-IDF transformation will not only normalize DTM, but also increase the weight of words/terms which appear in a single document/sentence or limited number of documents/sentences and inversely decrease the weight of words/terms used in a lot of documents.\par
	
	\item Creating a dictionary of sentiment bearing phrases and labeling a set of documents. In most cases in a document, there are adjective noun combinations like `horrible pitching' and `devastating loss.' These combinations are called \textit{sentiment bearing phrases}. By assigning a number to the sentiment, each sentiment bearing phrases get a relative score. These scores are predetermined by how frequently a given phrase occurs near a set of known positive words (e.g. good, wonderful, spectacular) and a set of negative words (e.g. bad, horrible, awful).\par
	
	Using an extremely large corpus of text (via an Internet search engine), the algorithm evaluates the nearness of known positive and negative words to the phrase being considered. Then we verify whether we should associate that phrase with positive or negative sentiment, and just how closely we should associate it. For each sentiment bearing phrase, search engine queries are used. Each query returns a hit count. These hit counts are combined using a mathematical operation called a log odds ratio to determine the score for a given phrase. The odds ratio is generally used in text mining to rank the works based on the relevancy of class by using the frequency of words \citep{schutze2008introduction}.\par
	The pseudo code for calculating the log odds ratio is as follows:  \par
	
	IF $hit\_positive\_count < hit\_negative\_count$ THEN \par
	\indent $ratio = -1*(hit\_negative\_count/hit\_positive\_count)$\par
	IF $hit\_positive\_count > hit\_negative\_count$ THEN \par
	\indent $ratio = hit\_positive\_count/hit\_negative\_count$\par
	ELSE $ratio = 0$ \par
	IF $ratio < -10$ THEN $ratio = -10$\par
	IF $ratio > 10$ THEN $ratio = 10$
	
	\item Training a sentiment classification model. Lasso and elastic-net regularized generalized linear models \citep{friedman2010regularization} are utilized to train the labeled documents in the previous step and the performance of the model is evaluated with 10-fold cross-validation. This is done using the \emph{glmnet} package in R.
	\item Scoring the sentences. All sentences are represented as predicted probabilities of positiveness using the model built in the previous step.
	
\end{enumerate}	

\subsection{Application: Transportation Concepts}
\begin{figure}
		\begin{minipage}{\linewidth}
			\includegraphics[width=1\linewidth]{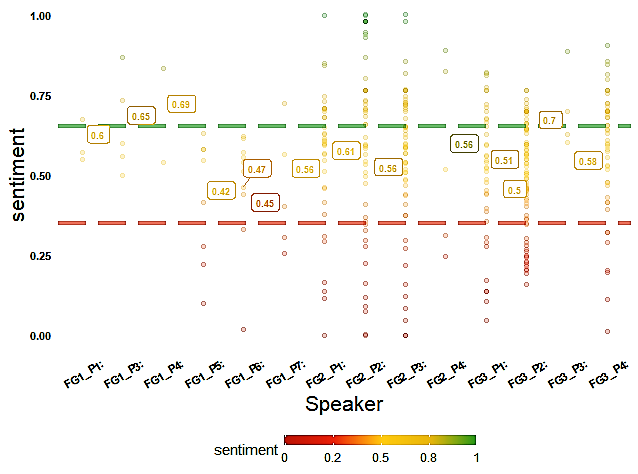} \hfill
			\includegraphics[width=1\linewidth]{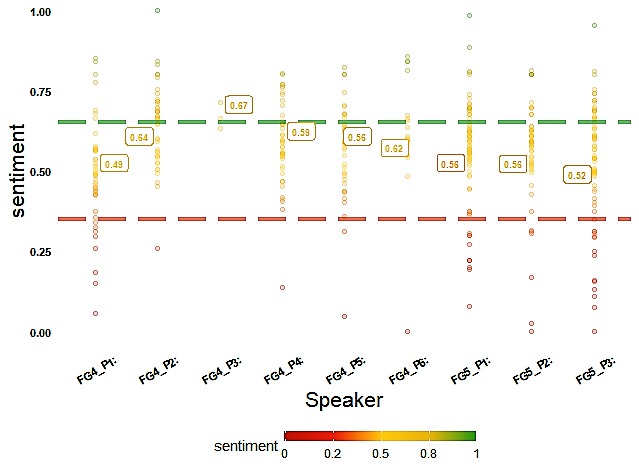}%
		\end{minipage}
	\caption{Average probability of positiveness for Humboldt Park (top) and Evanston (bottom) focus group participants discussing the \textbf{built environment}. The left end of the sentiment spectrum (red) represents more negative tones while the right end (green) represents more positive tones; the boundaries as shown by the dashed lines in the figure demarcate probability thresholds of 0.33 and 0.67, respectively. Each dot denotes an uninterrupted utterance by each speaker during the focus group discussion and the average sentiment across all utterances of a speaker is displayed in the accompanying box.}
	\label{fig:SA1}
\end{figure}

\begin{figure}
		\begin{minipage}{\linewidth}
			\includegraphics[width=1\linewidth]{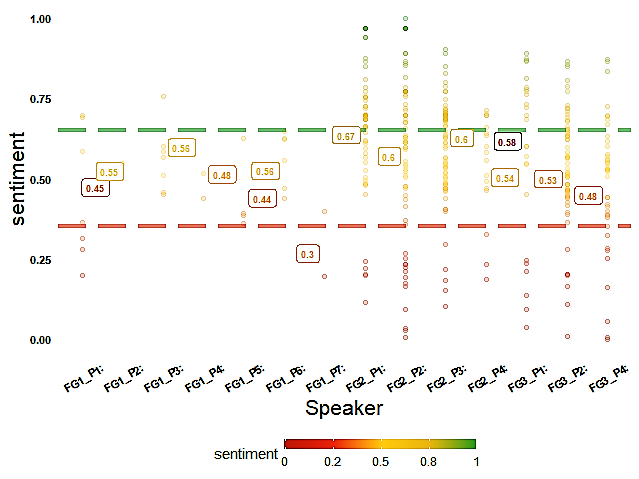} \hfill
			\includegraphics[width=1\linewidth]{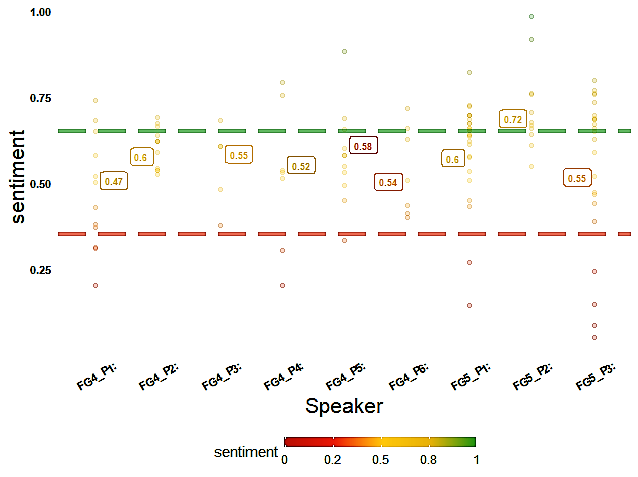}%
		\end{minipage}
	\caption{Average probability of positiveness for Humboldt Park (top) and Evanston (bottom) focus group participants discussing \textbf{well-being}. Remaining interpretation as shown in Figure 4}
	\label{fig:SA2}
\end{figure}

\begin{figure}
		\begin{minipage}{\linewidth}
			\includegraphics[width=1\linewidth]{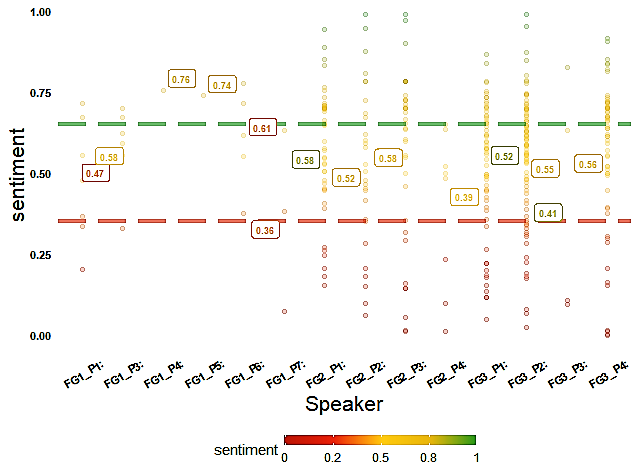} \hfill
			\includegraphics[width=1\linewidth]{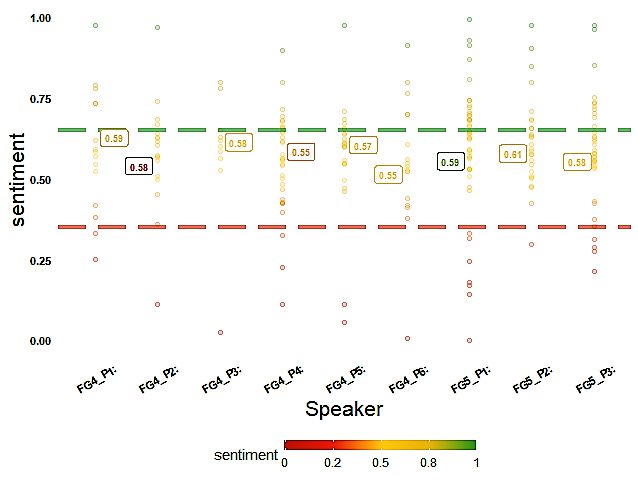}%
		\end{minipage}
	\caption{Average probability of positiveness for Humboldt Park (top) and Evanston (bottom) focus group participants discussing \textbf{culture and community}. Remaining interpretation as shown in Figure 4}
	\label{fig:SA3}
\end{figure}

To tease out overall differences between Humboldt Park (HP) and Evanston (EV), we separated the focus group dialogue based on the corresponding primary topic and applied the sentiment analysis procedure described in the previous subsection. Figures \ref{fig:SA1}-\ref{fig:SA3} illustrate the average \textit{probability of positiveness} for each individual in discussions on the (a) built environment, (b) well-being, and (c) cultural and community identities. The fourth topic, focusing on ideas for improvements in daily mobility, is excluded from this analysis due to its considerably shorter length compared to the other topics. The figures show the focus group participants on the x-axis and the positiveness probability on the y-axis. The red and green bars respectively represent the 0.35 and 0.65 thresholds discussed earlier. The graph labels are the corresponding averages for each speaker. Lengen and Kistemann (\citeyear{Lengen2012}) note that emotion is an important component of place identity expression, which further justifies this application of sentiment analysis. \par

We then took the mean probability of positiveness per topic, denoted by the following $\mu$-values, to show an aggregate sentiment trend: for [T1], $\mu_{HP}$ = 0.561 and $\mu_{EV}$ = .579; for [T2], $\mu_{HP}$ = 0.524 and $\mu_{EV}$ = .570; for [T3], $\mu_{HP}$ = 0.545 and $\mu_{EV}$ = .578. We see from these results that the Evanston participants were consistently more positive than Humboldt Park participants, with the biggest difference observed in the well-being discussion, which also has the lowest relative score for both community areas. Especially for Humboldt Park, this may reflect a desire for transport and public health policy that has a stronger emphasis on individuals' quality of life. \par

Finally, we examined the overall \textit{sentiment scores} (as opposed to probabilities) associated with different modes of transportation. We searched all phrases related to each mode and compiled a set of corresponding sentiment scores.  Results are displayed in Table 3. Several observations are worth noting:

\begin{enumerate}
	\item The private car is the only mode in both community areas to have a negative overall score, which we interpret to echo the expressed desire by participants to move away from automobile dependency as addressed in the topic modeling results. Hagman (\citeyear{Hagman2003}), however, points out that this does not easily translate into action because individuals tend to overlook their own contributions to automobile culture's negative consequences. Also, the relatively larger standard deviations compared to other modes have different meanings between the two community areas. For Humboldt Park, positive phrases highlighted the desire for traffic calming measures that would improve safety perceptions; for Evanston, positive phrases emphasized the leisure utility of car use, as well as having access to a private vehicle without having to rely on it to satisfy daily mobility needs.
	\item For the remaining four modes, Evanston participants were more positive than Humboldt Park participants in all cases except concerning public transportation. This phenomenon may arise from the fact that residents in Humboldt Park rely more heavily on public transportation, namely bus, and are thus more appreciative of its ability to provide access to resources that the community area lacks. Meanwhile, Evanston is growing in its multimodalism, so residents can afford to be somewhat more critical of transit operations while praising its function as the backbone of a sustainability-driven neighborhood.
	\item Expanding on this point, the most pronounced difference between Evanston and Humboldt Park is in the shared mobility + multimodal category. To address the first component, while both community areas have generally positive attitudes towards bikeshare and ridesourcing, Evanston residents possess the economic (e.g. credit cards) and technological (e.g. cell phones) resources to utilize the emerging mobility services. Meanwhile, Humboldt Park participants view these modes somewhat favorably, particularly bikeshare as a potential conduit of physically active lifestyles in addition to new job opportunities for younger residents graduating from bicycle maintenance training programs.
\end{enumerate}

\begin{center}
	\begin{threeparttable}
		\scriptsize
		\captionsetup{justification=centering}
		\caption[labelsep=period]{Sentiment Scores for Transport Mode Categories}
		\begin{tabular}{|>{\centering\arraybackslash}m{6cm}||>{\centering\arraybackslash}m{1.5cm}|>{\centering\arraybackslash}m{1.5cm}|>{\centering\arraybackslash}m{1cm}|@{}m{0pt}@{}} 
			\hline
			\textbf{Mode} & \textbf{Mean} & \textbf{Std. Dev.} & \textbf{Count} &\\ [5pt]
			\hline \hline
			\multicolumn{4}{|c|}{\textbf{Humboldt Park}} &\\
			\hline \hline
			Walking & 0.050 & 0.546 & 44 &\\ [5pt]
			\hline
			Bicycling & 0.113 & 0.446 & 69 &\\ [5pt]
			\hline
			Public Transportation (Bus, Train) & 0.168 & 0.432 & 52 &\\ [5pt]
			\hline
			Private Car & -0.168 & 0.514 & 27 &\\ [5pt]
			\hline
			Shared Mobility + Multimodal & 0.160 & 0.460 & 54 &\\ [5pt]
			\hline \hline	
			\multicolumn{4}{|c|}{\textbf{Evanston}} &\\
			\hline \hline
			Walking & 0.167 & 0.536 & 78 &\\ [5pt]
			\hline
			Bicycling & 0.207 & 0.519 & 77 &\\ [5pt]
			\hline
			Public Transportation (Bus, Train) & 0.110 & 0.531 & 93 &\\ [5pt]
			\hline
			Private Car & -0.089 & 0.571 & 70 &\\ [5pt]
			\hline
			Shared Mobility + Multimodal & 0.361 & 0.296 & 37 &\\ [5pt]
			\hline	
		\end{tabular}
	 \begin{tablenotes}
		\small
		\item Note: the mean score for each mode is calculated by taking the sum of all sentiment scores (ranging from -1 to 1) associated with a mention of that mode, divided by the number of occurrences in the text (indicated by the Count column).
	\end{tablenotes}
	\end{threeparttable}
\end{center}

\section{Discussion of policy propositions}
The fourth and final discussion topic represents a synthesis of what participants believe their community area's transportation priorities should be within the next five years. Resembling the findings from topic modeling and sentiment analysis, as well as those in a recent review of Dutch cycling policies \citep{Harms2015}, there is significant overlap between Humboldt Park and Evanston.
\begin{itemize}
	\item Invest in travel-based education, including on-demand trip planning via mobile phone apps and sharing space in a multimodal environment.
	\item Introduce traffic calming measures and protected bicycle lanes to improve safety, in addition to empowering women to feel secure when walking or taking public transportation.
	\item Construct a continuous bicycle network across the city, but avoid radial design that mimics the faults of urban highway systems; Heinen et al. (\citeyear{Heinen2010}) touch on the importance of network design in promoting cycling.
	\item Use social marketing \citep{Davies2012} to reduce the status-seeking behavior associated with car use as well as the `elitist' perceptions of cyclists.
	\item Create an `explanation of benefits' for active transportation use, highlighting individual and environmental health.
	\item Connect shared mobility modes, such as bikeshare, to community centers to engender a collectivist mindset.
\end{itemize}
Hence this list represents a set of shared societal interests that policymakers could utilize to address large-scale issues.

Humboldt Park participants, however, express three distinct concerns considering innovation in transportation. First, as a neighborhood of lower socioeconomic status, there is a need to work through a `hierarchy of needs' before introducing new mobility services. These basic conditions include universal Wi-Fi access, which is important for accessing navigation and information-seeking tools, and education in one's mobility options. Second, participants are wary of policies that could be associated with spurring gentrification. While some individuals were adamant in disassociating transportation investments with gentrification, others suggested that infrastructure improvements or improved mobility options will probably increase the neighborhood's attraction to developers and non-locals who are economically privileged. This may be construed as a threat to the cultural integrity of minority neighborhoods; the idealization of Chicago as a tapestry of separate but connected neighborhoods through an equitable transportation network reflects this notion. Third, autonomous vehicles, albeit briefly mentioned, were contextualized differently between the two community areas. Whereas Evanston participants mainly discussed them as a threat against multimodal travel environments, Humboldt Park participants stressed the importance of `autonomous individuals' rather than vehicles. To expound, there is an emerging fear that such technological advancement, particularly in the transit industry, will lead to severe job loss among low-income households, further widening socioeconomic gaps that innovation should aim to lessen. Therefore, the community area would benefit more, especially framed from a psychological well-being perspective, from improved public transportation services, since they are an established norm for daily mobility. One proposition would be to involve neighborhood members in the design of new bus stops and rail stations that would reflect local culture \citep{Rietveld2004} and boost self-esteem through a revitalized sense of ownership \citep{Davies2012}. Thus, local governments that promote such bottom-up approaches to transportation planning \citep{Marsden2017} could assuage concerns over gentrification as traditionally disadvantaged populations will feel empowered within their roles in fostering local multimodalism.  \par

\section{Conclusions}
Active mobility research has established that individual, social and spatial factors need to be considered to design effective interventions \citep{gotschi2017towards}. The current paper adds to the literature by studying the local community dimensions using mixed qualitative and quantitative methods. Examining the concept of active mobility within the narratives of two distinct community areas reveals key factors for achieving equitable and sustainable transport policy at different geographic scales. The novel application of topic modeling and sentiment analysis to focus group discourse highlights the need for additional research incorporating the infusion of qualitative data with emerging quantitative techniques, for the purposes of building up the repertoire of people-centric transportation planning practices. In a general sense, while residents of different neighborhoods are bound to share common goals for improving society, there will be area-specific factors that should not be overlooked, particularly for the advancement of traditionally-disadvantaged populations \citep{Lubin2016}. This research shows that, in the case of Humboldt Park and Evanston, there exists what one might call \textit{mobility tension}; to expound, as technological innovation and concerns for the environment and well-being continue to grow in a sometimes disharmonious manner, conflict between different groups of travelers arises within multimodal environments, propelled by the need for individuals to re-learn certain travel habits for the sake of collective compromise.

To help counter this tension, the ideas presented in this paper suggest that comparative analysis could assist policymakers in developing a framework for community areas to assess their `readiness for' specific transportation investments. As pointed out by Weber (\citeyear{Weber2014}), literature on bicycle and pedestrian policy implementation in the U.S. is scarce, so there is a pressing need to explore novel conceptual frameworks to assist transportation planning efforts. For instance, stage theories of behavior change represent a psychology-based market segmentation strategy that could illuminate successful pathways for implementing tailored interventions (see, for instance, Bamberg et al. (\citeyear{Bamberg2011})), but this approach often lacks a multilevel perspective. Creating a community-based stages-of-change framework that characterizes the acceptance of or resistance to emerging \textit{mobility cultures} would introduce ideas related to measuring (a) dissonance between an individual and his or her community area \citep{Schwanen2005} and (b) patterns of movement between stages over time to study the longitudinal effects of transportation innovation \citep{Kroesen2014}. Additionally, researchers should integrate metrics accounting for possible self-selection biases in order to disentangle built environment influences on travel behavior \citep{Mokhtarian2008}, although this might be a less severe problem when investigating populations exhibiting smaller degrees of social mobility. \par

Adopting a strategy in accordance with this line of thinking would require a data collection effort that is mindful of cultural theories stemming from sociological research. To demonstrate, Hannerz (\citeyear{hannerz_1992}) explores the processes constituting cultural formation in modern society based on the following three principles: (1) ideas and modes of thought that relates to underlying concepts and values, (2) forms of externalization that function as physical place-making representations of the collective cognitive state, and (3) the spatial and social distributions of these fundamental elements of culture. In research that attempts to characterize populations in a homogeneous manner due to shared geopolitical boundaries (such as for community areas in the current study), it is critical for researchers to address the possibility of interacting \textit{subcommunities} whose differences regarding the first two principles could induce tensions, due to perceived inequities in policy treatment. Therefore, although physical built environments and local knowledge conveyed through social networks might allow a certain degree of neighborhood-scale generalization when it comes to matters of transportation, there is bound to be significant heterogeneity in individual psychosocial attributes that must be addressed properly through any quantitative analysis. Returning to the idea of market segmentation, which has been explored in previous qualitative research on travel behavior \citep{Nielsen2015}, one possible direction would be to characterize subcommunities within a stage-of-change framework based on sociodemographic, life stage, and lifestyle information \citep{Haustein2013}, which could serve as input variables to latent class cluster analysis \citep{McDonald2012}. The overall research goal would be to capture heterogeneity across the stages regarding symbolic and affective meaning associated with new forms of mobility in the context of cultural integrity and preservation.

Our research study is not without its limitations, though. Foremost among these is that focus group participants were recruited via convenience sampling through established networks rooted in the research team composition. As with most focus group work, representation is limited, although in our case the balance between men and women was precisely even across the whole sample. In Humboldt Park, focus group participants were all of Hispanic heritage and admittedly more informed on the topics than the average resident. Thus, it is erroneous to assume that Black residents, who comprise the majority of the population in the western half of the community area, and individuals who are not involved with educational programming originating from the Paseo Buricua district share the same perspectives. As for Evanston, participants were middle-aged and elderly white people; hence the discussion does not necessarily reflect the broader socio-demographic composition within the city limits, particularly members of the Millennial generation and racial/ethnic minority groups. Even though the findings are not truly generalizable, the utility of our novel methodological approach should be explored further in future research studies, particularly those that utilize larger sets of documents to improve the reliability of text mining results. Furthermore, the interpretation of findings still involves some subjectivity, although we argue it is less prevalent than in traditional content analysis. That being said, research that invites subjective interpretation is vital to promoting healthy dialogue between policymakers and neighborhood residents, especially when equity is a central issue. \par

In summary, this comparative analysis of two community areas in the Chicago region illuminates an ideology of reduced car dependence that appears to be fairly independent of socioeconomic status. The process of implementing policies for achieving low-carbon mobilities, especially walking and cycling behaviors, should acknowledge differences in built environment perceptions, well-being concerns, and identities that comprise local cultures and communities. As a result, achieving a higher quality of life through diverse and inclusive transportation planning is more attainable.

\section*{Acknowledgments}
The research team would like to thank the following individuals and institutions for their assistance in making this study possible: the staff of Humboldt Park's Puerto Rican Cultural Center and Diabetes Empowerment Center, the Age Friendly and Citizens' Greener Evanston task forces, and Katherine Knapp, who served as Evanston's Transportation and Mobility Coordinator when the focus groups were conducted. This research was supported by the Junior William A. Patterson
Professor Chair in Transportation at Northwestern University of author Stathopoulos. Partial support was given by the NU Institute for Complex Systems (NICO) Data Science Initiative award titled ?Is ours better than mine: Investigating collaborative consumption systems? for authors Biehl, Chen and Stathopoulos.

\section*{Appendix}
Discussion Guide for Focus Groups
\begin{enumerate}
\item [T1] Built environment
	\begin{enumerate}
	\item	What are your perceptions about the local transportation network? What about other elements of the built environment?
	\item What services, resources, and amenities are accessible to residents within HP/EV? Is there anything important missing and are residents able to easily access this elsewhere?
	\item Do residents tend to compare HP/EV with other community areas in Chicago? Does this lead to a feeling that there is something that HP/EV comparatively possesses or lacks?
	\end{enumerate}
\item [T2] Well-being
	\begin{enumerate}
	\item Describe the overall physical and psychological well-being of HP/EV residents. Are there programs in place to encourage healthier lifestyles?
	\item Is the provided definition of \textit{subjective well-being} satisfactory? What changes would you make if you were trying to measure an individual's happiness?
	\item It is possible that encouraging changes in travel behavior or the provision of new transportation infrastructure could impact well-being. Could you think of 			specific connections that may exist?
	\end{enumerate}
\item [T3] Cultural and community identities
	\begin{enumerate}
	\item Describe the social and cultural settings of HP/EV. What are notable norms and values that define groups within the local community?
	\item How could discussions of the built environment and community identity relate to one another? Is there a prevailing `sense of community' in HP/EV or is there a noteworthy divergence in interests?
	\item Does the local transportation system work for everyone? Should future transport and planning policy attempt to be more inclusive of certain groups of people?
	\end{enumerate}
\item [T4] Improvements
	\begin{enumerate}
	\item What should be goals for HP/EV within the next five years regarding transportation and improved mobility for residents?
	\end{enumerate}
\end{enumerate}

\bibliographystyle{tfcad}
\bibliography{Mobilities_ref,focus}

\end{document}